\documentclass[12pt]{article}

\usepackage[a4paper,margin=1in]{geometry}  
\usepackage{graphicx}
\usepackage{floatrow}
\usepackage{epstopdf}
\usepackage{float}
\usepackage{booktabs}
\usepackage{braket}
\usepackage{amsmath,amssymb}      
\usepackage{algorithm}
\usepackage{algpseudocode}
\usepackage{authblk}  
\usepackage{setspace} 
\usepackage{titlesec} 
\usepackage{abstract} 
\usepackage{hyperref}

\titleformat{\section}{\normalfont\large\bfseries}{\thesection.}{1em}{}
\titleformat{\subsection}{\normalfont\normalsize\bfseries}{\thesubsection.}{1em}{}

\title{\textbf{The Role of Connection Density in an Adaptive Network with Chaotic Units}}

\author[1]{Ramiro Pl\"{u}ss}
\author[1]{Pablo Martín Gleiser}
\affil[1]{\normalsize Instituto Tecnológico de Buenos Aires, CABA C1437, Argentina}
\date{} 

\begin{document}
\maketitle

\begin{abstract}
We investigate the role of connection density in an adaptive network model of chaotic units that dynamically rewire based on their internal states and local coherence. By systematically varying the network’s connectivity density, we uncover distinct dynamical regimes and structural transitions, revealing mechanisms of spontaneous modularity, dynamical segregation, and integration. We find that at higher densities, the network exhibits both local clustering and global synchrony. Additionally, we observe that low-density networks tend to fragment into desynchronized clusters, while high-density networks converge to synchronized states combining strong global integration with persistent modular segregation. Inspired by neural architectures, this model provides a general framework for understanding how simple microscopic rules can give rise to complex emergent behaviors in dynamical networks.
\end{abstract}

\maketitle

\section{\label{sec:Introduction}Introduction:}

Adaptive networks in which the topology co-evolves with the state of dynamical units are a powerful framework for studying emergent complexity in nonlinear systems. The interplay between structure and function can lead to rich spatiotemporal patterns, including synchronization, modularity, and spontaneous reorganization \cite{kaneko1989, kaneko1994, gong2004,gleiser2006,gomez2009,haimovici2013,cascallares2015}. These systems are often inspired by neural architectures, where the balance between functional segregation and integration serves as a fundamental organizational principle \cite{sporns2013, deco2015}. Segregation refers to the specialization of local brain regions into functionally distinct modules, typically characterized by dense intra-modular connectivity. In contrast, integration reflects the capacity of distant brain areas to coordinate and communicate effectively, relying on inter-modular or long-range connections to support coherent global dynamics. Disruptions in this balance have been implicated in a variety of neurological and cognitive conditions. For example, altered states of consciousness, including deep sleep \cite{spoormaker2010, tagliazucchi2014}, anesthesia, and coma \cite{boly2017}, have been associated with changes in network segregation and integration. Likewise, neurodegenerative diseases such as Alzheimer’s \cite{stam2007} and psychiatric disorders such as schizophrenia \cite{griffa2015, chang2019} have been linked to abnormal network dynamics.\\

In this work, we study how the density of connections in an adaptive network model modulates its emergent properties. Specifically, we extend a network model of coupled chaotic maps with state-dependent rewiring, originally proposed by Gong and Van Leeuwen \cite{gong2004}, by introducing the network's link density as an additional control parameter. By varying the density of connections and the strength of interactions between units, we observe the emergence of distinct organizational patterns, including fragmentation at low densities, local clustering reflecting segregation at intermediate densities, and modular integration combining segregation and integration at higher densities. Each configuration is associated with characteristic patterns of synchronization and coherence. Our findings reveal that connection density plays a central role in shaping the network's self-organization. Notably, we observe non-monotonic relationships between interaction strength and clustering, as well as transitions between regimes of high segregation and high integration. These structural features include the balance between modular specialization (functional segregation), global coordination (functional integration), and the dynamic reconfiguration of network states \cite{sporns2013, deco2015}.

\section{\label{sec:Model}Model:}

The model considered in this work \cite{gong2004} starts with the construction of an undirected random network comprising \( N = 300 \) nodes. Random networks are generated according to the Erd\H{o}s--R\'enyi \( G(N, p) \) model, where each possible edge between nodes is independently included with probability \( p \), following a Bernoulli distribution \cite{erdds1959}. This yields an expected number of links \( L_c = p \cdot \binom{N}{2} = p \cdot \frac{N(N-1)}{2} \), and an expected average degree \( \braket{k} = p \cdot (N - 1) \). We used \( p = \lbrace 0.03,\ 0.06,\ 0.12,\ 0.18 \rbrace \), which correspond to expected values \( \braket{k} \approx \lbrace 9,\ 18,\ 36,\ 54 \rbrace \), and \( L_c \approx \lbrace 1350,\ 2700,\ 5400,\ 8100 \rbrace \), respectively. The network dynamics are governed by coupled logistic maps, where each node evolves according to the function \( f(x) = 1 - a x^2 \), with \( a = 1.7 \), a parameter value chosen to induce chaotic behavior. At time step zero, each node \( i \) is initialized with a random activity \( x_i^0 \sim \mathcal{U}(-1, 1) \) \( \forall i \in \{0, \ldots, N-1\} \). The state at time \( n \), denoted \( x_i^n \), evolves iteratively. The set of neighbors of node \( i \) is denoted by \( B(i) \), and \( M(i) \) indicates the number of such neighbors. The system is evolved for 300{,}000 time steps, and the update rule for computing the state \( x_i^{n+1} \) at time step \( n+1 \) is given by Eq.~(\ref{eq:coupled_logistic_map}).

\begin{equation}
x^{i}_{n+1} = (1-\varepsilon)f(x^{i}_{n},a) + \frac{\varepsilon}{M_i} \sum_{j,j \in B(i)} f(x^{j}_{n},a)\label{eq:coupled_logistic_map}
\end{equation}

The neighbors of unit $i$ are the units that have a direct connection with unit $i$. The network remains undirected. The parameter \( \varepsilon \) denotes the coupling strength. In the $n$-th time step, the coherence between unit $i$ and unit $j$, $d_{ij}(n)$, is defined as the absolute value of the difference between the activation values of 
the units, as in Eq.~(\ref{eq:coherence}).

\begin{equation}
d_{ij}(n) = |x^{i}_{n} - x^{j}_{n}|\label{eq:coherence}
\end{equation}

To implement adaptive rewiring, a temporal window $T$ is defined as the number of iterations between successive topology updates. Every $T$ steps, each node evaluates its average coherence with neighbors and rewires by replacing its least coherent connection with a more coherent non-neighbor. In this work, $T = 800$. The system is iterated up to 300{,}000 time steps, during which each unit adaptively rewires its connections based on the dynamic coherence previously described. The network is reconnected according to the dynamic coherence between its units. Since the total number of links $L_c$ is directly related to the average degree $\langle k \rangle$. We use this measure of connection density to remain consistent with existing literature and report results in terms of $\langle k \rangle$, as the number of nodes $N$ is fixed and the total number of links depends directly on the average degree.

\subsection{\label{sec:Algorithm}Algorithm:}

Starting from the initial random structure, the following algorithm is applied. $(I)$ A random value for the activity or state of each node 
is chosen from the range $(-1, 1)$ for the entire system. $(II)$ The state of the system is calculated based on Eq.~(\ref{eq:coupled_logistic_map}),
and discard an initial transient time $T$. $(III)$ A random node $i$ is selected, and the value of $d_{ij}(T + 1)$ is 
computed with Eq.~(\ref{eq:coherence}) for all nodes in the network. Two nodes, $j_1$ and $j_2$, are identified based on their \( d_{i,j}(T + 1) \) values. 
Specifically, $j_1$ corresponds to the node with the minimum $d_{i,j}(T + 1)$ among all nodes, while $j_2$ is selected from 
the neighbors of $i$ and has the maximum $d_{i,j}(T + 1)$ value. $(IV)$ If $j_1$ is one of the neighbors of $i$, no changes are 
made to the connection. However, if $j_1$ is not a neighbor, the connection between units $i$ and $j_2$ is replaced by the 
connection between nodes $i$ and $j_1$. $(V)$ The algorithm returns to the Step $(I)$ and repeats these steps.

\subsection{\label{sec:Network Metrics}Network Metrics:}

To quantify the emergent structure of the network, we computed standard topological measures at each iteration of the simulation. We denote by \( N \) the total number of nodes in the network. The local clustering coefficient of a node \( i \), denoted \( c_i \), is defined as \( c_i = 2e_i / [k_i(k_i - 1)] \), where \( k_i \) is the degree of node \( i \), and \( e_i \) is the number of existing links among its neighbors. The global clustering coefficient \( C \) is then obtained by averaging \( c_i \) over all nodes in the network: \( C = (1/N) \sum_{i=1}^{N} c_i \). To evaluate global integration, we compute the average shortest path length \( L \), defined as \( L = \sum_{i \ne j} d(i, j) / [N(N - 1)] \), where \( d(i, j) \) is the shortest path between nodes \( i \) and \( j \), restricted to the largest connected component. Finally, we compute the small-world index \( \omega \), which captures the simultaneous presence of high local clustering and short global path lengths. It is defined as \( \omega = (C / C_{\text{rand}}) / (L / L_{\text{rand}}) \), where \( C \) and \( L \) are the clustering and path length of the actual network, and \( C_{\text{rand}} \) and \( L_{\text{rand}} \) are the corresponding values computed over Erdős–Rényi random graphs with the same number of nodes and expected average degree. For the time-resolved metrics \( C \), \( L \), and \( \omega \), we computed the corresponding network property every 1{,}000 time steps over a total duration of 300{,}000 time steps, yielding 300 time points per simulation. For each parameter configuration, we ran 10 independent simulations and computed the mean and standard deviation across those realizations at each sampled time point. In contrast, the value of \( \langle C \rangle \) reported in the figures represents a final-state measure. For each individual simulation, we computed the average of the global clustering coefficient \( C \) over the last 100 sampled time points, which correspond to the final 100{,}000 time steps of the simulation. These values provide a representative estimate of the system's stationary behavior. The resulting averages were computed across the 10 independent simulations corresponding to the same combination of coupling strength \( \varepsilon \) and mean degree \( \langle k \rangle \). Therefore, \( \langle C \rangle \) corresponds to the average across simulations of the global clustering coefficient in the final stage of the dynamics.\\

To further characterize the modular organization of the network, we applied community detection to each network snapshot using the Louvain algorithm \cite{blondel2008}, which iteratively maximizes the modularity of the partition to identify cohesive communities within the network. This yields a partition of the network into non-overlapping clusters, or communities, denoted by \( \{\mathcal{S}_1, \mathcal{S}_2, \ldots, \mathcal{S}_{n_s}\} \), where \( \mathcal{S}_i \) is the set of nodes assigned to cluster \( i \), and \( n_s \) is the total number of detected clusters at a given time step. From this partition, we defined the following quantities: the size of the largest cluster \( S_1 = \max_i |\mathcal{S}_i| \), the size of the second-largest cluster \( S_2 = \text{second-largest } |\mathcal{S}_i| \), the mean cluster size \( \langle s \rangle = \frac{1}{n_s} \sum_{i=1}^{n_s} |\mathcal{S}_i| \), and its standard deviation \( \sigma_s = \sqrt{ \frac{1}{n_s} \sum_{i=1}^{n_s} \left( |\mathcal{S}_i| - \langle s \rangle \right)^2 } \). These quantities provide complementary information: \( S_1 \) captures the presence of a dominant module, \( S_2 \) helps detect secondary structures or fragmentation, and \( \langle s \rangle \) together with \( \sigma_s \) describe the heterogeneity of community sizes.\\

We computed normalized histograms representing the empirical distributions of network metrics, such as the degree distribution, ensuring that the total area under each curve sums to 1.

\section{\label{sec:Results}Results:}

Building on the work of Gong and Van Leeuwen~\cite{gong2004}, who identified dynamical regimes as a function of $\varepsilon$ for a fixed network density, we extend this adaptive network model of chaotic units by systematically exploring how different network densities $\langle k \rangle$ modulate these transitions and structural outcomes. We identified three distinct regimes depending on the connection density $\langle k \rangle$ and the coupling strength $\varepsilon$: (i) for low densities regardless of the coupling strength, the network tends to fragment and become disconnected regardless of the coupling value, due to insufficient connectivity. Thus, at intermediate and high densities, we observe two possible behaviors: (ii) for low coupling, the network preserves a randomly structured topology over time; and (iii) for sufficiently large coupling values, the system undergoes a transition to a small-world structure. To better understand these behaviors, we separately analyzed the structural metrics associated with each regime. We used connection probabilities $p = \lbrace 0.03,\ 0.06,\ 0.12,\ 0.18 \rbrace$, which correspond to expected average degrees $\braket{k} \approx \lbrace 9,\ 18,\ 36,\ 54 \rbrace$ for $N = 300$. Accordingly, the expected number of links is $L_c \approx \lbrace 1350,\ 2700,\ 5400,\ 8100 \rbrace$, respectively. For $\langle k \rangle = 9$, we observed partial network disconnection. Due to fragmentation, many nodes become isolated, resulting in substantial variation in the number of connected nodes across simulations. Consequently, structural metrics such as global clustering and average shortest path length are no longer comparable across conditions, since they refer to networks with fundamentally different connectivity properties, particularly in terms of the number of connected nodes. We focus our analysis on the cases $\langle k \rangle = \{18, 36, 54\}$, where the network does not fragment. Although the initial network size is \( N = 300 \), the adaptive rewiring rule can lead to the removal of nodes that become completely disconnected. In practice, node loss is rare for networks with average degrees \( \langle k \rangle \geq 18 \), particularly under moderate to strong coupling strengths. However, for low connection densities such as, \( \langle k \rangle = 9 \) or weak coupling strength \( \varepsilon \), fragmentation is more common and may result in the elimination of isolated nodes.

\subsection{\label{sec:Emergence of Small-World Properties in the Network Structure}Emergence of Small-World Properties in the Network Structure:}

\begin{figure}[h]
\begin{center}
\includegraphics[scale=0.2]{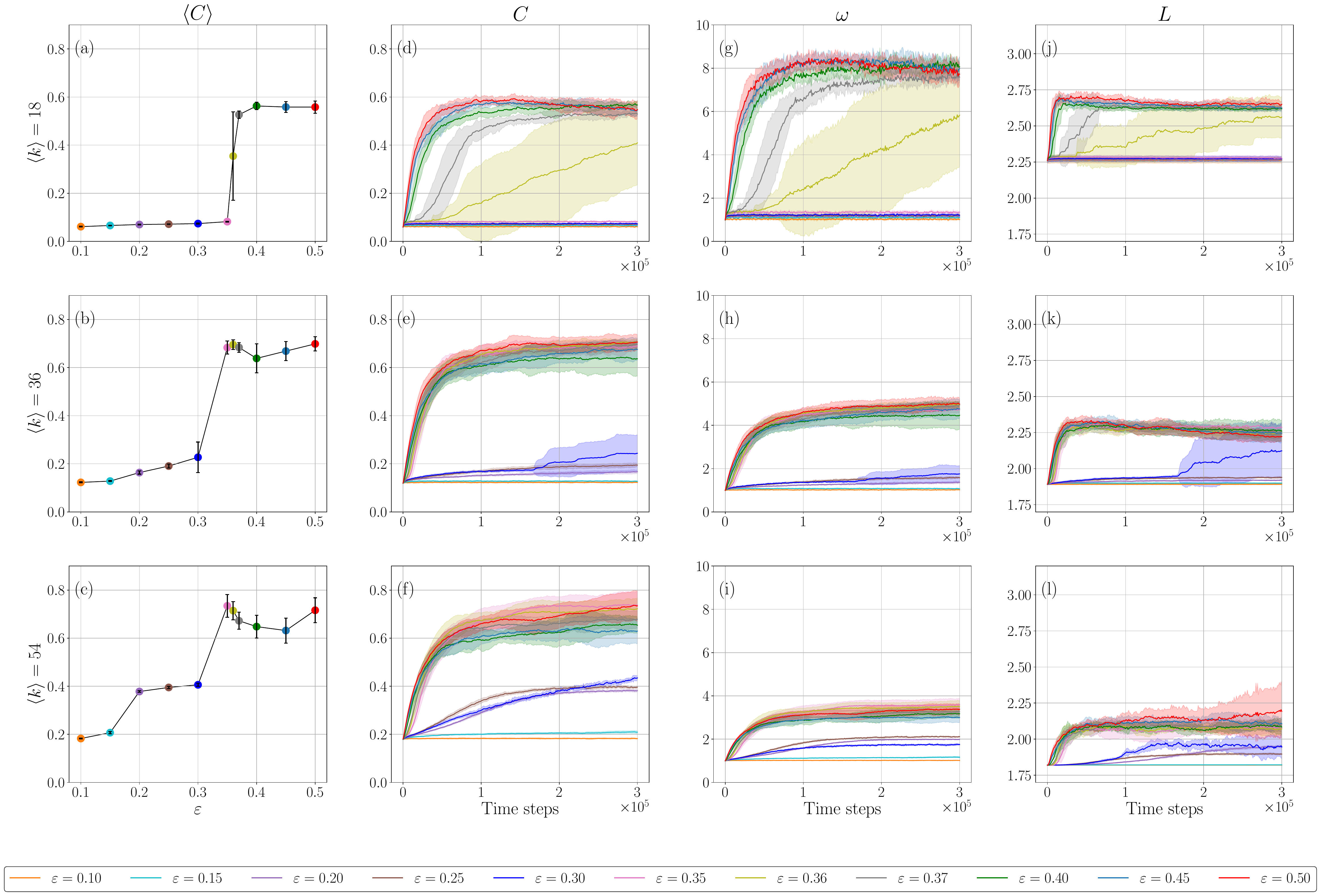}
\end{center}
\caption{(a–c) Average clustering coefficient $\langle C \rangle$, temporal evolution of (d–f) global clustering coefficient $C$, (g–i) small-world index $\omega$, and (j–l) average shortest path length $L$, for networks with $\langle k \rangle = 18$ (top row), $\langle k \rangle = 36$ (middle row), and $\langle k \rangle = 54$ (bottom row). All metrics are computed across adaptive rewiring events, with shaded regions indicating the standard deviation across realizations.}
\label{fig:clswi}
\end{figure}

To characterize the structural properties of the network, we focus primarily on the average clustering coefficient ($\langle C \rangle$). We also tracked the temporal evolution of the global clustering coefficient ($C$), the small-world index ($\omega$), and the average shortest path length ($L$) to understand how these features evolve over time \cite{watts1998,strogatz2001}. Fig~(\ref{fig:clswi}) summarizes these analyses across networks with $\langle k \rangle = \{18, 36, 54\}$ and coupling strengths $\varepsilon \in \{0.10, 0.15, 0.20, 0.25, 0.30, 0.35, 0.36, 0.37, 0.40, 0.45, 0.50\}$. We observe that a distinct small-world regime emerges, marked by high $\langle C \rangle$, $C$ and $\omega$. In contrast, $L$ shows limited sensitivity to $\varepsilon$ and structural changes. Notably, $\omega$ increases with $\varepsilon$ across all $\langle k \rangle$, but its peak value decreases with connectivity, being highest for $\langle k \rangle = 18$.

The clustering coefficient $C$ emerges as the primary driver of $\omega$ variations. Around critical coupling values, strong dispersion across realizations signals structural transitions. Data points are sampled every $1{,}000$ time steps, and shaded regions represent the standard deviation across multiple realizations, capturing variability in the dynamics. Asymmetry around the mean in some regimes reflects structural transitions or competing configurations. By comparing the average clustering coefficient $\langle C \rangle$ across different network densities $\langle k \rangle = \lbrace 18, 36, 54\rbrace$ and coupling strengths $\varepsilon$ in Fig.~(\ref{fig:clswi}a–c), we observe that $\langle C \rangle$ generally increases with $\varepsilon$, with sharp transitions marking the onset of modular organization. For $\langle k \rangle = 18$, a clear transition occurs at $\varepsilon^* \approx 0.36$, separating random-like from small-world regimes (Fig.~(\ref{fig:clswi}a,d)). For $\langle k \rangle = 36$, the transition appears earlier at $\varepsilon^* \approx 0.30$ in Fig.(~\ref{fig:clswi}b,e), while for $\langle k \rangle = 54$, it shifts further left, below $\varepsilon^* < 0.20$ (Fig.~\ref{fig:clswi}c,f), indicating faster reorganization but diminishing small-world effects at high density. Notably, Fig.~(\ref{fig:clswi}a–c) also shows that the trend of $\langle C \rangle$ shifts from monotonic to non-monotonic as $\langle k \rangle$ increases, reflecting the complex interplay between density and dynamical reconfiguration. In Fig.\ref{fig:clswi}j–l, $L$ shows limited sensitivity to variations in $\varepsilon$ and to the underlying structural changes. Given that the small-world index $\omega$ primarily depends on both $C$ and $L$, we can focus our analysis on $C$ and, consequently, on $\langle C \rangle$. Nevertheless, as shown in Fig.~(\ref{fig:clswi}g–i), $\omega$ increases with $\varepsilon$ across all $\langle k \rangle$, but its peak value decreases with increasing connectivity, being highest for $\langle k \rangle = 18$ and lowest for $\langle k \rangle = 54$.

\subsection{Interplay Between Structure and Emergent Dynamics}

\begin{figure}[h]
\centering
\includegraphics[scale=0.07]{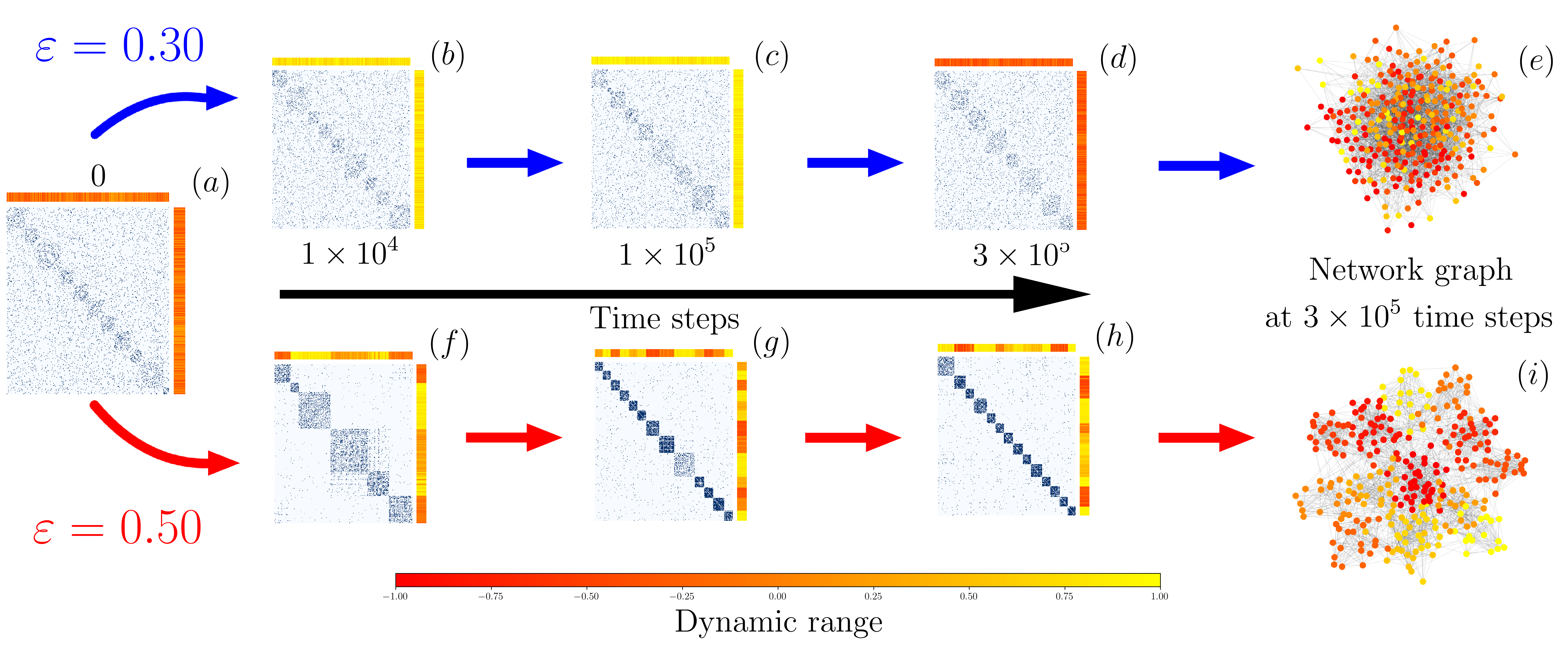}
\caption{Temporal evolution of community structure and node dynamics for $\langle k \rangle = 18$ over $3\times10^5$ time steps. (a) Shows the initial random network. (b–d) Illustrate the evolution under $\varepsilon = 0.30$. (f–h) Under $\varepsilon = 0.50$, at selected time steps (0, $1\times10^4$, $1\times10^5$, and $3\times10^5$ time steps). Panels (e) and (i) display the final network graphs, with node colors encoding the dynamical states.}
\label{fig:louvain_k_18}
\end{figure}

\begin{figure}[h]
\centering
\includegraphics[scale=0.15]{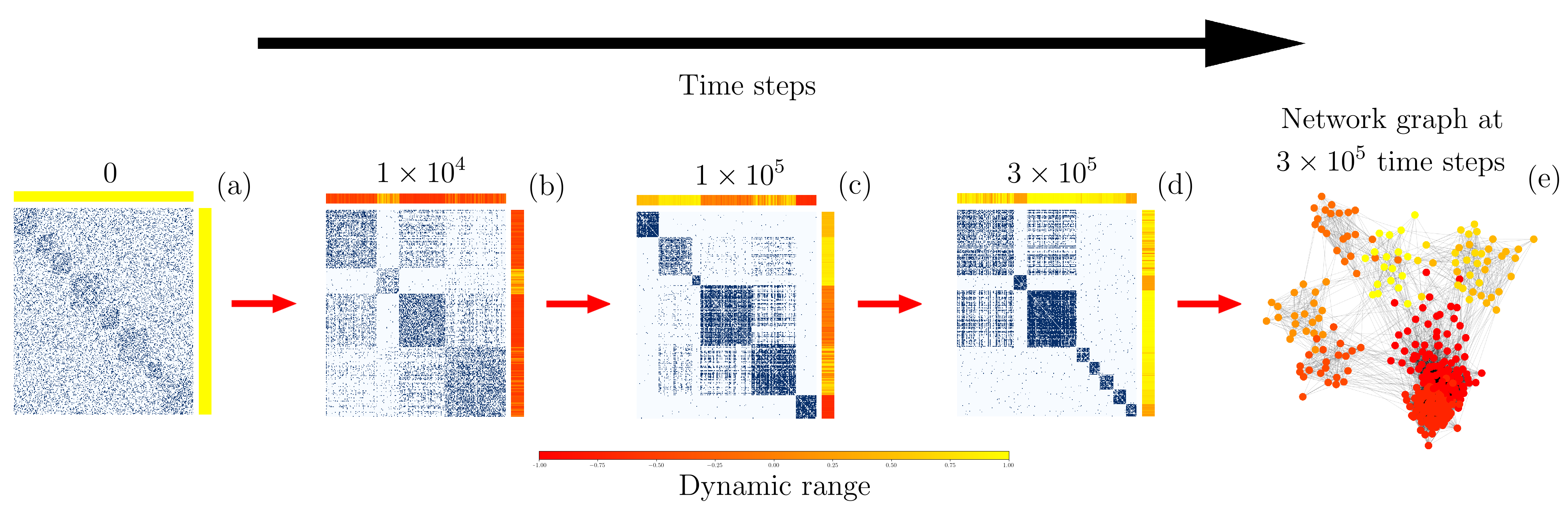}
\caption{Temporal evolution of community structure and node dynamics for $\langle k \rangle = 54$ over $3\times10^5$ time steps.(a–d) Show adjacency matrices at selected time steps (0, $1\times10^4$, $1\times10^5$, and $3\times10^5$ time steps) under $\varepsilon = 0.50$. Panel (e) display the final network graph, with node colors encoding the dynamical states.}
\label{fig:louvain_k_54}
\end{figure}

To visualize the interplay between structure and dynamics, we construct a figure where each panel represents the network's adjacency matrix, reordered according to the modular structure detected by the Louvain community detection algorithm. Color bars alongside the matrices indicate the dynamical states of the nodes, ordered consistently with the structural partitioning. At the same time, the color bar below the matrices reflects the dynamic range of the node states at each time step. This enables the visualization of emergent patterns such as synchronization blocks and community structures based on dynamical coherence. Subsequently, we analyze the modular organization using both qualitative and quantitative approaches across varying $\langle k \rangle$ and $\varepsilon$ values. Fig.~(\ref{fig:louvain_k_18}) illustrates the evolution of the network structure and node states at different time steps, obtained using the Louvain community detection method~\cite{blondel2008}, for $\varepsilon = \{0.30, 0.50\}$ and $\langle k \rangle = 18$. Fig.~(\ref{fig:louvain_k_18}a) shows the initial random configuration, serving as a common starting point. For weak coupling ($\varepsilon = 0.30$; Fig.~(\ref{fig:louvain_k_18}b–d)), no clear modular organization emerges even after prolonged evolution. The color bars reveal no evident synchronization patterns at early stages (Fig.~(\ref{fig:louvain_k_18}b,c)), although partial synchronization develops by the final state (Fig.~(\ref{fig:louvain_k_18}d)). However, this occurs without significant structural modularity, as confirmed in the final network layout (Fig.~(\ref{fig:louvain_k_18}e)). In contrast, for strong coupling ($\varepsilon = 0.50$; Fig.~(\ref{fig:louvain_k_18}f-h)), a progressive emergence of modular structure is observed, driven by synchronization. Initially (Fig.~(\ref{fig:louvain_k_18}f)), communities begin to form, often dominated by nodes with similar dynamical states, clustering around $x_n \approx 1.0$ (yellow) and $x_n \approx -1.0$ (orange). Although a fully stable configuration is not yet achieved, coherence within clusters becomes increasingly evident. As the evolution proceeds Fig.~(\ref{fig:louvain_k_18}g-h), the large initial communities gradually fragment into smaller, dynamically coherent subcommunities, reflecting a refinement of modular organization. The final network layout Fig.~(\ref{fig:louvain_k_18}i) clearly shows these structurally and dynamically coherent modules. For $\langle k \rangle = 54$, as shown in Fig.~(\ref{fig:louvain_k_54}), the increased number of links results in a denser adjacency matrix, leading to a more homogeneous connectivity pattern that initially obscures modular organization. Nonetheless, pronounced modular blocks emerge along the diagonal, indicating strong intra-community connectivity, as well as noticeable off-diagonal structures, particularly in Fig.~(\ref{fig:louvain_k_54}c,d). These off-diagonal blocks reflect integration between different communities, suggesting that nodes across distinct modules share similar dynamical states. This phenomenon is further illustrated in the final network layout in Fig.~(\ref{fig:louvain_k_54}e), where inter-community coherence is visually apparent.

To support and complement the qualitative observations from the modularity-reordered adjacency matrices, we further analyzed how coupling strength and average degree influence the emergence, consolidation, and fragmentation of communities by tracking the evolution of key cluster metrics: the sizes of the largest clusters ($S_1$, $S_2$), the average cluster size, and the total number of clusters. For $\langle k \rangle = 18$, weak coupling ($\varepsilon = 0.30$) preserves structural stability over time, while strong coupling ($\varepsilon = 0.50$) induces a progressive fragmentation into smaller clusters. At $\langle k \rangle = 36$, the system initially remains stable, but under weak coupling it exhibits delayed structural reorganization around iteration $1.5\times10^5$, characterized by large fluctuations in cluster sizes and number, suggesting a critical-like transition. For $\langle k \rangle = 54$, strong coupling leads to the rapid formation of stable, well-organized communities early in the simulation, whereas weak coupling results in a more gradual, fragmented reorganization with an increasing number of small clusters. Overall, these observations highlight the interplay between coupling strength and network density in shaping the modular organization and stability of the system over time.

\begin{figure}[h]
\centering
\includegraphics[scale=0.4]{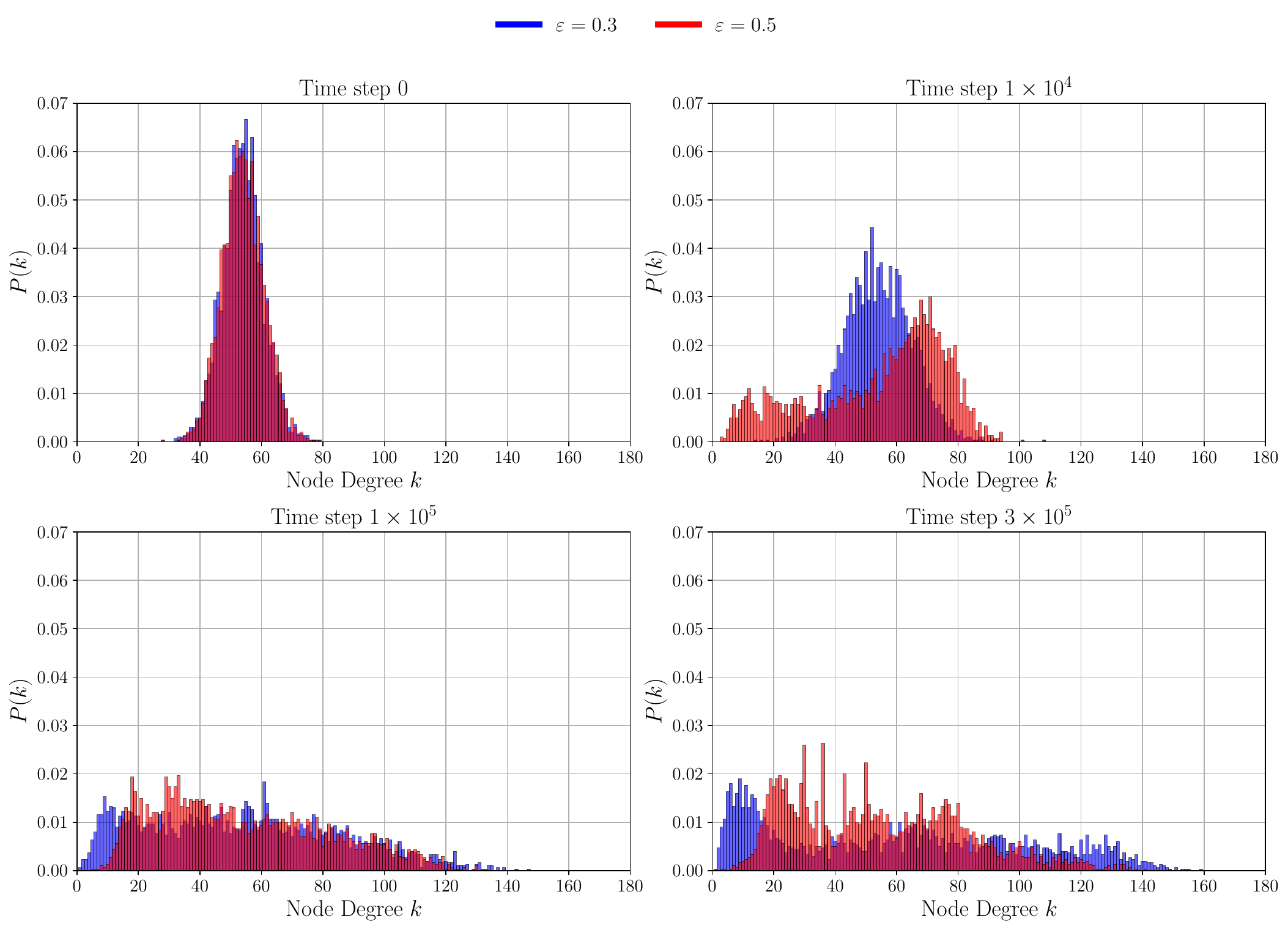}
\caption{Degree distribution \( P(k) \) at different time steps for \(\langle k \rangle = 54\). 
(a) Initial state (0), 
(b) early state (\(1 \times 10^4\)), 
(c) intermediate state (\(1 \times 10^5\)), and 
(d) final state (\(3 \times 10^5\)). 
Each panel shows the degree distribution, comparing \(\varepsilon = 0.3\) (blue) and \(\varepsilon = 0.5\) (red).}
\label{fig:temporal_evolution_k54}
\end{figure}

Additionally, to gain further insights into the evolving local structure, we examined the degree distribution \( P(k) \) across different initial average degrees \(\langle k \rangle = \{18, 36, 54\}\). For \(\langle k \rangle = \{18, 36\}\), the distributions remained relatively narrow and centered, exhibiting only minor shifts or slight asymmetries during network evolution. 
However, for \(\langle k \rangle = 54\) in Fig.~(\ref{fig:temporal_evolution_k54}), the distribution showed a marked broadening and signs of multimodality, particularly under strong coupling (\(\varepsilon = 0.50\)), where the formation of hubs and a more heterogeneous connectivity pattern became evident. These results suggest that high initial connectivity amplifies the impact of adaptive rewiring on the network's local topological organization.\\

\section{Conclusions}

Our results extend previous investigations into adaptive rewiring and chaotic dynamics \cite{gong2004} by systematically exploring how variations in average network density \(\langle k \rangle\) influence topological transitions. We show that increasing \(\langle k \rangle\) alters the strength of small-world characteristics and reshapes the balance between segregation and integration in the emergent network structures. The model reveals distinct structural regimes. At low densities (\(\langle k \rangle = 9\)), we observe network fragmentation. At intermediate densities (\(\langle k \rangle = \{18, 36\}\)), two regimes emerge: a random regime at low coupling \(\varepsilon\), and a small-world regime with high clustering (as a measure of segregation) at higher \(\varepsilon\), where well-defined modules appear. Increasing the average degree \(\langle k \rangle\) lowers the critical coupling \(\varepsilon^*\) required for the onset of modular organization. For \(\langle k \rangle = 36\), we observe a non-monotonic increase in \(\langle C \rangle\). This suggests that excessive density can blur modular boundaries, leading to partial desynchronization between communities and a reduction in small-world properties. In both intermediate-density cases, the network exhibits clear segregation but lacks sufficient integration. Importantly, the interplay between topology and dynamics captured by this model is consistent with empirical findings in structural connectivity between healthy controls and patients with schizophrenia \cite{griffa2013,griffa2015}. Thus, the intermediate-density configuration reflects findings in schizophrenia, where brain organization shows altered modularity and deficient global integration, particularly affecting control and default mode networks. At higher densities (\(\langle k \rangle = 54\)), we again observe both dynamical regimes, with similar non-monotonic trends in clustering measures and a further decline in small-worldness. However, in this regime, segregation and integration coexist, partly due to the emergence of hubs that bridge distinct communities, thereby promoting global integration by shortening the paths between modules. The appearance of off-diagonal structures in the adjacency matrix reflects connections between distinct modules, indicating the presence of integration across segregated communities. This is further supported by the presence of hubs, as revealed in the degree distribution histograms. This emergent architecture reflects the hierarchical organization of the human brain, where structural hubs such as the precuneus or thalamus facilitate coordination across functional networks. In contrast, multiple studies have shown that in schizophrenia, hubs exhibit both structural and functional alterations, impairing the brain's ability to integrate distributed information. In this sense, this model not only reproduces transitions across different connectivity regimes but also illustrates how the emergence or disruption of hubs may be associated with functional states ranging from healthy to pathological. The observed results highlight the delicate balance between segregation and integration: stronger coupling promotes intra-community synchronization but may impair global clusterization when the network becomes overly dense.\\

\section*{Data Availability and Reproducibility}

All simulations were implemented in Python 3.11. The implementation relies on the following open-source libraries: \texttt{NumPy 1.26}, \texttt{NetworkX 3.2.1}, \texttt{pandas 2.2.1}, \texttt{matplotlib 3.8.3}, \texttt{python-louvain 0.16} for community detection, and \texttt{Numba 0.60} for just-in-time compilation of performance-critical functions. All simulation parameters are specified in a separate configuration file (\texttt{config.json}), and the code is structured for modularity and extensibility. The entire simulation pipeline can be reproduced by running the main script with \texttt{python main.py}. A \texttt{requirements.txt} file is provided to facilitate environment replication. The simulation outputs (e.g., adjacency matrices, node states, and computed metrics) are stored in a structured directory under \texttt{results/}, which is automatically generated by the code. These outputs include all data used to produce the figures presented in this work. The data that support the findings of this study are openly available in  
\textcolor{blue}{\texttt{\href{https://github.com/ramirop2021/The-Role-of-Connection-Density-in-an-Adaptive-Network-with-Chaotic-Units}{GitHub}}}.\\

\bibliographystyle{unsrt} 

\end{document}